\documentclass[conference]{IEEEtran}
\IEEEoverridecommandlockouts

\usepackage[T1]{fontenc}
\usepackage{cite}
\usepackage{amsmath}

\usepackage{newtxtext,newtxmath}

\usepackage{algorithm}
\usepackage{algorithmic}
\usepackage{graphicx}
\usepackage{textcomp}
\usepackage{xcolor}
\usepackage{booktabs}
\usepackage{url}
\usepackage[hidelinks]{hyperref}
\usepackage{orcidlink}
\usepackage{tikz}
\usetikzlibrary{
  arrows.meta,
  shapes.geometric,
  shapes.symbols,
  positioning,
  decorations.pathreplacing,
  calc
}
\usepackage{pgfplots}
\pgfplotsset{compat=1.18}
\definecolor{siline}{HTML}{4A3AA7}
\definecolor{cusumline}{HTML}{EB6834}
\definecolor{refgray}{HTML}{898781}
\usepackage{stfloats}
\usepackage{balance}

\makeatletter
\providecommand{\bstctlcite}[1]{\@bsphack
  \@for\@citeb:=#1\do{%
    \edef\@citeb{\expandafter\@firstofone\@citeb}%
    \if@filesw\immediate\write\@auxout{\string\citation{\@citeb}}\fi}%
  \@esphack}
\makeatother

\newcommand{\DR}{52.0\%}          
\newcommand{\FAR}{20.8\%}         
\newcommand{\AUC}{0.851}          
\newcommand{\Lmean}{1.8\,s}       
\newcommand{\Nwindows}{770}       
\newcommand{\Npos}{102}           
\newcommand{\Nneg}{668}           
\newcommand{\CUSUMh}{1.0}         
\def\BibTeX{{\rm B\kern-.05em{\sc i\kern-.025em b}\kern-.08em
    T\kern-.1667em\lower.7ex\hbox{E}\kern-.125emX}}

\begin{document}
\addtolength{\textheight}{2\baselineskip}

\bstctlcite{IEEEexample:BSTcontrol}

\title{Detection of Synchronized AI Data Center Load Episodes Using SCADA Telemetry}

\author{%
\IEEEauthorblockN{%
Chandan~Chaudhary\orcidlink{0009-0002-2389-9568},
Abanish~Tiwari\orcidlink{0009-0003-0609-8571},~\emph{Student Member, IEEE},
Yansong~Pei\orcidlink{0000-0002-4647-7491},~\emph{Member, IEEE},\\
Mohammed~Ben-Idris\orcidlink{0000-0002-8731-8913},~\emph{Senior Member, IEEE},
and Joydeep~Mitra\orcidlink{0000-0001-9287-0983},~\emph{Fellow, IEEE}%
}
\IEEEauthorblockA{%
Electrical and Computer Engineering, Michigan State University, East Lansing, MI 48824, USA\\
E-mails: chaud152@msu.edu;\; tiwariab@msu.edu;\; peiyanso@msu.edu;\; benidris@msu.edu;\; mitraj@msu.edu%
}
\vspace{-2.25em}
}

\maketitle

\begin{abstract}
AI data centers running distributed training workloads impose episodic, spatially correlated active-power disturbances on the transmission grid. These synchronized episodes increase cross-substation load correlation and limit the diversification benefit that reserve margin planning assumes. Conventional energy management systems evaluate each substation independently and do not extract the cross-substation statistical structure that defines a synchronized episode. This paper develops a detection method that identifies synchronized AI data-center load episodes from standard active-power telemetry without new instrumentation, trained classifiers, or labeled data. The method computes a Synchronization Index, the dominant eigenvalue fraction of a sliding sample covariance matrix across substations. A cumulative-sum (CUSUM) sequential test converts the index into a delay-bounded episode alarm. The same eigendecomposition yields, at no additional cost, a dominant eigenvector that attributes a detected episode to the substations that drive it. Tests on a real-time digital simulator (RTDS) model of the IEEE 39-bus system with three AI data-center buses show that the method separates episode and normal windows with a wide margin over chance and attributes episode participation at substation granularity.
\end{abstract}


\vspace{-0.75em}

\section{Introduction}
Global data center electricity demand reached approximately 415\,TWh in 2024 and is projected to more than double by 2030~\cite{iea2024electricity,epri2024powering,cheng2026ai}. Individual facilities now enter interconnection queues at 500\,MW to 2\,GW~\cite{nerc2025lltf,epri2024powering}, and approximately 70\% of U.S. data center capacity is concentrated in a handful of electricity market regions, with Northern Virginia as the most prominent example~\cite{nerc2025lltf}. Prior work shows that this geographic concentration produces correlated active-power patterns across multiple transmission substations when co-located facilities share a distributed AI training workload~\cite{chaudhary2026spatial,chaudhary2026adequacy,chaudhary2027gridedge}. The choice of data-center load model also changes predicted grid stability outcomes. Dynamic converter-based representations show larger frequency excursions than static constant-power or ZIP models on the same test system~\cite{chaudhary2025stability}. As AI training clusters grow in scale and geographic density, real-time detection of these synchronized load episodes becomes essential, and existing grid monitoring infrastructure cannot provide it.

A distributed AI training job synchronizes graphics processing unit (GPU) nodes across multiple campuses through a common orchestration platform, and its periodic communication phase drives a correlated active-power fluctuation across multiple transmission buses~\cite{chaudhary2026predispatch,chaudhary2027gridedge,chaudhary2026semimarkov}. Field measurements confirm that large-scale training jobs induce fast, synchronized transients at repetition rates within Supervisory Control and Data Acquisition (SCADA) observability~\cite{li2024unseen,choukse2025power,go2025characterizing,narayanan2021megatron,jimenez2025datacenter,li2025aidynamics}.

The Energy Management System (EMS) cannot see synchronized AI loads as synchronized. Every EMS alarm function operates on a single measured quantity at a single location~\cite{wood2014power}. A substation active-power channel trips an alarm when its reading exceeds a threshold but has no view of whether neighboring substations are moving in a correlated pattern. An operator who monitors three data center-connected buses sees three independent alarm panels. A synchronized episode is therefore indistinguishable from three coincidental load changes, although the reserve adequacy implication differs fundamentally. The measurements exist at one-second SCADA resolution, yet standard EMS alarm functions do not extract the cross-substation statistical structure that defines a synchronized episode. The North American Electric Reliability Corporation (NERC) has flagged this coincident behavior as an emerging reliability risk with no adequate monitoring solution~\cite{nerc2025lltf,takci2025flexibility}.

Load-characterization studies confirm that synchronized episodes exist and quantify their spatial structure~\cite{li2024unseen,choukse2025power,chaudhary2026spatial,chaudhary2026modal,wang2016proactive,zhang2025mitigating}, but each relies on retrospective analysis of historical records and therefore does not provide a real-time detector. Wide-area eigenvalue and principal-component monitoring methods target generator-side oscillations and frequency events at phasor measurement unit (PMU) timescales and are not designed for correlated-load detection at SCADA resolution~\cite{phadke2008wams,kwon2025operational,rafferty2016pca}. Classical change-point methods provide scalar process alarms, and sequential detection theory has been applied to line-outage and cyber-attack detection in power systems~\cite{page1954cusum,basseville1993detection,chen2016outage,rovatsos2017statistical,kurt2018distributed}, yet these single-variable formulations do not extend to the multi-bus active-power correlation problem. A companion effort detects correlated AI facility loads in real time on substation-deployable edge hardware, but only as a binary classification between independent and correlated operation, without a continuous severity index or a spatial attribution diagnostic~\cite{chaudhary2027gridedge}. None of these bodies of work attributes a detected episode to the substations that drive it.

Existing methods therefore leave two gaps unresolved. Real-time, continuous severity scoring of a synchronized episode from standard SCADA telemetry is absent, and spatial attribution that identifies which substations drive a detected episode remains undeveloped. The contributions of this paper close both gaps in turn.

\begin{enumerate}

\item It introduces the \emph{Synchronization Index} (SI), the dominant eigenvalue fraction of a sliding sample covariance matrix over data center-connected substations. Its independence baseline of $1/N$ is analytically known and requires no labeled training data. A Cumulative Sum (CUSUM) sequential test converts the index stream into a delay-bounded alarm whose threshold is calibrated over an episode-free startup period.

\item It establishes that the dominant eigenvector of the same covariance estimate is a spatial participation vector that identifies which substations drive the synchronized mode. This attribution costs nothing beyond the eigendecomposition already required for the Synchronization Index.

\end{enumerate}

The remainder of this paper is organized as follows. Section~\ref{sec:episodes} characterizes the synchronized episode phenomenon. Section~\ref{sec:method} develops the full detection and attribution pipeline. Section~\ref{sec:results} presents the RTDS testbed and detection results along with the factors that bound their validity. Section~\ref{sec:conclusion} synthesizes these findings and outlines future research directions.


\section{Synchronized AI Load Episodes}\label{sec:episodes}

This section establishes the physical mechanism, observable SCADA signature, and statistical characterization of synchronized AI data center load episodes.

\subsection{Physical Mechanism}

\begin{figure}[!t]
  \centering
  \includegraphics[width=\columnwidth, trim={4cm 10.9cm 10.85cm 5.45cm},clip]{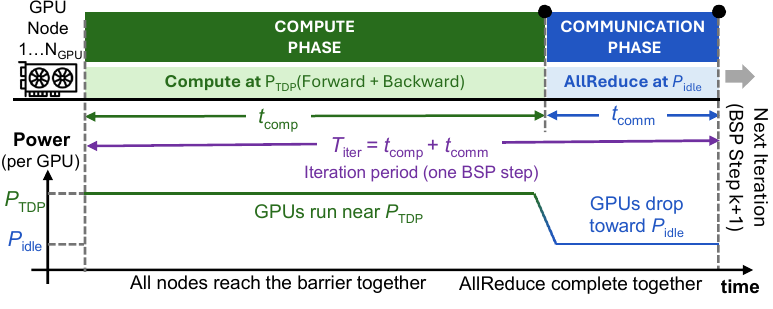}
  \vspace{-0.8cm}
  \caption{GPU power-phase behavior under one BSP training iteration \protect \cite{chaudhary2026predispatch}.}
  \vspace{-1.25em}
  \label{fig:gpu}
\end{figure}

\begin{figure}[!htbp]
  \centering
     \vspace{-1.25em}
  \resizebox{\columnwidth}{!}{%

\begin{tikzpicture}[scale=0.8, every node/.style={scale=0.8}]

\begin{scope}[shift={(-0.3,0)}]

\node[font=\scriptsize, text=red!70!black]  at (3.5,5.8) {Transmission Level};
\node[font=\scriptsize, text=blue!70!black, align=center] at (3.5,3.55) {Subtransmission\\Level};

\node[circle,draw,thick,minimum size=0.8cm,fill=red!12] (B1) at (0,5.0)   {\small Bus 1};
\node[circle,draw,thick,minimum size=0.8cm,fill=red!12] (B2) at (3.5,5.0) {\small Bus 2};
\node[circle,draw,thick,minimum size=0.8cm,fill=red!12] (B3) at (7.0,5.0) {\small Bus 3};

\draw[very thick,red!60!black](B1)--(B2) node[midway,above,font=\scriptsize]{230 kV};
\draw[very thick,red!60!black](B2)--(B3) node[midway,above,font=\scriptsize]{230 kV};

\foreach \x in {0, 2.5, 4.5, 7.0}{
  \node[circle,draw,thick,minimum size=0.25cm,fill=white] at (\x,4.0){};
  \draw[thick](\x,3.8)--(\x,4.2);
  \draw[thick](\x-0.18,4.0)--(\x+0.18,4.0);
}
\draw[thick,blue!60!black](B1)--(0,4.0);
\draw[thick,blue!60!black](B2)--(2.5,4.0);
\draw[thick,blue!60!black](B2)--(4.5,4.0);
\draw[thick,blue!60!black](B3)--(7.0,4.0);

\node[circle,draw,thick,fill=blue!18,minimum size=0.8cm] (S1) at (0,2.8)   {\small Sub.1};
\node[circle,draw,thick,fill=blue!18,minimum size=0.8cm] (S2) at (2.5,2.8) {\small Sub.2};
\node[circle,draw,thick,fill=blue!18,minimum size=0.8cm] (S3) at (4.5,2.8) {\small Sub.3};
\node[circle,draw,thick,fill=blue!18,minimum size=0.8cm] (S4) at (7.0,2.8) {\small Sub.4};

\draw[thick,blue!60!black](0,4.0)  -- (S1);
\draw[thick,blue!60!black](2.5,4.0)-- (S2);
\draw[thick,blue!60!black](4.5,4.0)-- (S3);
\draw[thick,blue!60!black](7.0,4.0)-- (S4);
\draw[thick,blue!50](S1)--(S2) node[midway,above,font=\scriptsize]{69--138 kV};
\draw[thick,blue!50](S3)--(S4) node[midway,above,font=\scriptsize]{69--138 kV};

\node[rectangle,draw,thick,fill=orange!20,minimum width=1.9cm,minimum height=1.0cm,align=center]
    (DC1) at (0,0.7)
    {\scriptsize\textbf{DC1}\\[-1pt]\scriptsize 150 MW\\[-1pt]\scriptsize Training};
\node[rectangle,draw,thick,fill=orange!20,minimum width=1.9cm,minimum height=1.0cm,align=center]
    (DC2) at (2.5,0.7)
    {\scriptsize\textbf{DC2}\\[-1pt]\scriptsize 180 MW\\[-1pt]\scriptsize Training};
\node[rectangle,draw,thick,fill=green!15,minimum width=1.7cm,minimum height=0.9cm,align=center]
    (Load) at (4.5,0.7)
    {\scriptsize Independent\\[-1pt]\scriptsize Load 50 MW};
\node[rectangle,draw,thick,fill=orange!20,minimum width=1.9cm,minimum height=1.0cm,align=center]
    (DC3) at (7.0,0.7)
    {\scriptsize\textbf{DC3}\\[-1pt]\scriptsize 120 MW\\[-1pt]\scriptsize Inference};

\draw[->,thick,orange!70!black](DC1)  -- (S1)  node[midway,left, xshift=-2pt,font=\scriptsize]{$P_1$};
\draw[->,thick,orange!70!black](DC2)  -- (S2)  node[midway,right,xshift=2pt, font=\scriptsize]{$P_2$};
\draw[->,thick,orange!70!black](DC3)  -- (S4)  node[midway,left,             font=\scriptsize]{$P_3$};
\draw[->,thick,green!60!black] (Load) -- (S3)  node[midway,right,            font=\scriptsize]{$P_4$};

\coordinate (DC2_bot40) at ($(DC2.south west)!0.40!(DC2.south east)$);
\coordinate (DC2_bot60) at ($(DC2.south west)!0.60!(DC2.south east)$);

\draw[<->,thick,red!65,dashed]
    (DC1.south) to[bend right=20]
    node[midway,below=4pt,font=\scriptsize,text=red!70!black]{Sync'd Workload, Weather}
    (DC2_bot40);

\draw[<->,thick,red!65,dashed]
    (DC2_bot60) to[bend right=20]
    node[midway,below=1.5pt,font=\scriptsize,text=red!70!black]{Correlated}
    (DC3.south);



\node[font=\tiny,text=blue!70,anchor=south west] at (S1.north east) {\scriptsize SCADA};
\node[font=\tiny,text=blue!70,anchor=south east] at (S4.north west) {\scriptsize SCADA};

\fill[blue!45] (1.25,2.8) circle (2.5pt);


\node[draw,thick,rounded corners=3pt,fill=blue!6,
      minimum width=6.0cm,minimum height=0.72cm,align=center,font=\scriptsize]
    (PRE) at (3.35,-1.3)
    {\textbf{Preprocessing:}\;Time Sync $\to$ Detrend $\to$ Normalize};

\node[draw,thick,rounded corners=3pt,fill=blue!14,
      minimum width=6.0cm,minimum height=0.72cm,align=center,font=\scriptsize]
    (COV) at (3.35,-2.6)
    {\textbf{Sliding Covariance:}\;$\hat{\mathbf{C}}(t_k)$, win.\,$W$ $\to$ Eigendecomp.};

\node[draw,thick,rounded corners=3pt,fill=green!10,
      minimum width=6.0cm,minimum height=0.72cm,align=center,font=\scriptsize]
    (EMS) at (3.35,-3.9)
    {\textbf{EMS Detection:}\;SI\;|\;CUSUM\;|\;Participation};

\draw[->,thick,black!70] (PRE.south) -- (COV.north);
\draw[->,thick,black!70] (COV.south) -- (EMS.north);

\coordinate (PRE_inL) at ($(PRE.north)+(-0.3,0)$);
\coordinate (PRE_inR) at ($(PRE.north)+(0.3,0)$);

\draw[->,thick,dashed,blue!50]
    (S1.west) -- (-1.05,2.8) -- (-1.05,-1.3) -- (PRE.west);

\draw[->,thick,dashed,blue!50]
    (1.25,2.8) -- (1.25,-0.6) to[out=270,in=120] (PRE_inL);

\draw[->,thick,dashed,blue!50]
    (S4.east) -- (8.05,2.8) -- (8.05,-1.3) -- (PRE.east);

\fill[blue!45] (5.75,2.8) circle (2.5pt);
\draw[->,thick,dashed,blue!50]
    (5.75,2.8) -- (5.75,-0.6) to[out=270,in=60] (PRE_inR);

\node[font=\footnotesize,text=blue!55,rotate=90] at (-1.20,0.8)
    {SCADA telemetry (1\,s)};

\end{scope}

\begin{scope}[shift={(9,0)},scale=0.70]

\node[align=center,font=\scriptsize] at (-0.75,5.0) {%
$\mathbf{C} \!=\! \hspace{-0.15em}
\left[
\begin{array}{@{}cccc@{}}
1 & C_{12} & C_{13} & 0 \\
C_{12} & 1 & C_{23} & 0 \\
C_{13} & C_{23} & 1 & 0 \\
0 & 0 & 0 & 1
\end{array}
\right]
$};

\node[draw,thick,fill=red!10,
      minimum width=2.0cm,minimum height=2.25cm]
      (inset) at (0,1.2) {};

\node[font=\scriptsize,text=black] at (0,2.4) {Internal GPU Sync};

\node[rectangle,draw,fill=orange!30,minimum width=0.9cm,minimum height=0.2cm]
    at (-0.68,1.8)  {\tiny R$_1$};
\node[rectangle,draw,fill=orange!30,minimum width=0.9cm,minimum height=0.2cm]
    at (-0.68,1.15) {\tiny R$_2$};
\node at (-0.65,0.75) {\scriptsize$\cdots$};
\node[rectangle,draw,fill=orange!30,minimum width=0.9cm,minimum height=0.2cm]
    at (-0.68,0.35) {\tiny R$_n$};

\node[rectangle,draw,fill=cyan!30,minimum width=0.7cm,minimum height=0.32cm]
    at (0.75,1.20) {\tiny HVAC};

\node[font=\tiny,text=red!80!black] at (0,-0.15) {No spatial correlation};
\node[font=\tiny] at (0.55,0.40) {R\,=\,Rack};

\end{scope}

\draw[->,thick,blue!60!black]
    (DC3.east) -- ++(0.33,0.1);

\end{tikzpicture}  
}
  \caption{Conceptual overview of multiple AI data-center campuses sharing a common workload orchestration platform.}
  \vspace{-1.25em}
  \label{fig:concept}
\end{figure}
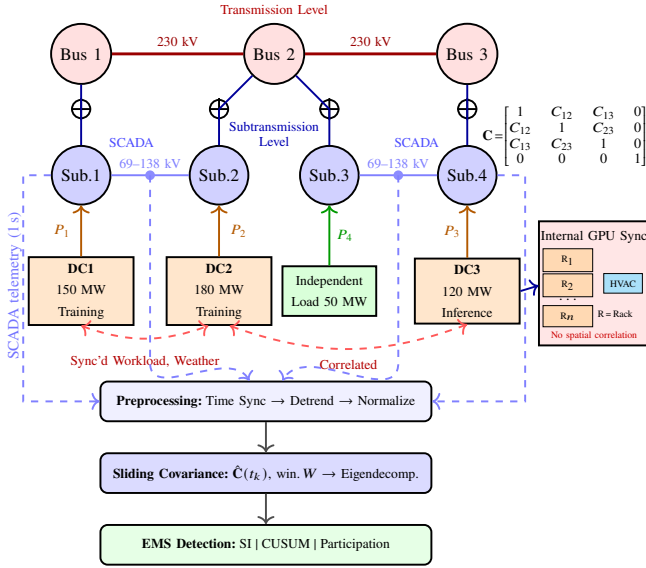

A synchronized AI load episode originates in the computational structure of large-scale distributed training. Modern deep learning models are trained using the bulk-synchronous parallel (BSP) paradigm, in which every training iteration is divided into two sequential phases~\cite{chaudhary2026predispatch, valiant1990bridging, chaudhary2026semimarkov}. In the \emph{compute phase}, GPU nodes process assigned data batches independently at high and nearly constant power draw. In the \emph{communication phase}, each node exchanges gradient updates with all other nodes before the next iteration can begin. This synchronization barrier draws a brief but large power spike as the inter-node communication fabric is saturated, and power then relaxes as the barrier clears and the next compute phase begins~\cite{go2025characterizing,choukse2025power}. Because the communication barrier is a global synchronization point, it affects every participating node at nearly the same instant, regardless of location. Fig.~\ref{fig:gpu} shows GPU power-phase behavior under one BSP iteration. A hierarchical semi-Markov model resolves this same BSP cycle into five power-distinguishable states and couples it to a facility-scale job-scheduling process that reproduces whole-facility power statistics~\cite{chaudhary2026semimarkov}.
When the GPU nodes of a single training job are distributed across multiple data-center campuses connected to separate transmission substations, the power profile of every facility reflects the same BSP iteration clock, as illustrated in Fig.~\ref{fig:concept}~\cite{chaudhary2026spatial}. The orchestration platform that dispatches the training job provides the common signal, and the grid observes the consequence~\cite{li2024unseen,li2025aidynamics}.

\subsection{Observable SCADA Signature}

From the perspective of SCADA active-power telemetry, a synchronized episode has three distinguishing characteristics. First, the active-power time series at multiple data center-connected substations transition simultaneously from a low-variance, approximately independent regime to a high-variance, correlated regime \cite{chaudhary2026spatial}. Second, the detrended active-power records at participating substations move together in the same direction and with similar shape because a common signal drives the shared motion \cite{chaudhary2026spatial}. Third, the episode persists for the duration of the gradient-synchronization barrier, typically 10--30\,s, before the load returns to independent compute-phase operation \cite{chaudhary2026predispatch}.

These characteristics distinguish a synchronized episode from two common confounders. An ordinary load coincidence produces simultaneous changes at multiple buses without waveform resemblance, and the detrended records move in unrelated directions. A single large-load event produces a magnitude change at one substation with no corresponding change at neighbors, whereas a synchronized episode requires simultaneous, directionally correlated fluctuations across multiple substations. That relational signature is its defining feature.

\subsection{Why Single-Bus Alarming Fails}

Conventional EMS threshold alarming cannot detect this pattern because it processes each substation channel independently. During a communication-phase spike, the per-substation active-power change may be large for a single facility yet remain within that substation's own alarm threshold, since the episode is defined not by any one bus reaching a dangerous level but by multiple buses moving together. A threshold check on bus $i$ asks only whether $i$ exceeds its own limit, never whether $i$, $j$, and $k$ move in correlated fashion. As a result, three simultaneous within-limit readings can occur at the instant they are most dangerous for reserve adequacy, while the alarm system reports nothing. This gap is a structural limitation of single-variable alarming, not a calibration problem a lower threshold would fix.

\subsection{Statistical Characterization}

The observable consequence of the mechanism described above is a change in the joint statistical structure of the multi-bus active-power vector. Let $\mathbf{p}(t) \in \mathbb{R}^N$ collect the detrended active-power measurements at $N$ data center-connected substations at time $t$. Under normal, independent operation, the loads at different substations are driven by independent local factors, and the covariance matrix $\mathbf{C} = \mathrm{E}[\mathbf{p}(t)\mathbf{p}(t)^\top]$ is approximately diagonal \cite{chaudhary2026spatial}. The matrix has full rank $N$, and its eigenvalues are roughly equal. This distribution indicates that no single direction dominates the variance in the data. During a synchronized episode, a common orchestration signal $s(t)$ enters every substation through a facility-specific weight $a_i$, so the active-power vector acquires a rank-one component,
\begin{equation}
  \mathbf{p}(t) = \mathbf{a}\,s(t) + \boldsymbol{\varepsilon}(t),
  \label{eq:signal_model}
\end{equation}
where $\mathbf{a} = [a_1,\ldots,a_N]^\top$ is the participation vector, $s(t)$ is the common synchronized signal with variance $\sigma_s^2 = \mathrm{E}[s^2(t)]$, and $\boldsymbol{\varepsilon}(t)$ is an independent residual with diagonal covariance $\mathbf{D}$, consistent with the spatial correlation structure measured directly in AI data-center-dominated power systems~\cite{chaudhary2026spatial}. The episode covariance becomes,
\begin{equation}
  \mathbf{C}_1 = \sigma_s^2\,\mathbf{a}\mathbf{a}^\top + \mathbf{D},
  \label{eq:episode_covariance}
\end{equation}
where $\sigma_s^2\mathbf{a}\mathbf{a}^\top$ is the synchronized common component shared across substations and $\mathbf{D}$ represents independent local noise, the rank-one-plus-diagonal structure of a spiked covariance model~\cite{Johnstone2001distribution}. The aggregate active power $P_\mathrm{agg}(t) = \mathbf{1}^\top\mathbf{p}(t)$ reveals the consequence of this shift. Under independence, the aggregate variance $\mathrm{Var}(P_\mathrm{agg})$ sums $N$ comparable per-bus variances and its standard deviation grows with $\sqrt{N}$, the diversification benefit reserve margin planning assumes, whereas under $\mathbf{C}_1$ the synchronized term scales the aggregate variance with $N^2$, so the standard deviation instead grows linearly with $N$~\cite{chaudhary2026adequacy}. Thus, the episode covariance $\mathbf{C}_1$ is rank-one plus diagonal. Nearly all of the covariance is concentrated in the direction of $\mathbf{a}$. The dominant eigenvalue therefore rises sharply, and the remaining eigenvalues represent only independent noise. This structural collapse from full-rank to approximately rank-one covariance is the signature the detector targets. This property of the joint distribution emerges precisely when the aggregate demand variability has grown most dangerous. This common-signal formalization matches the state-synchronized coupling introduced as a structural extension for facility-scale semi-Markov load models~\cite{chaudhary2026semimarkov}. In that extension, a dependent facility follows a reference facility's state transitions with a fixed probability instead of evolving independently.

\section{Detection Method}\label{sec:method}

This section develops a detection method that converts the covariance structural change identified for synchronized load episodes into a real-time severity index, a delay-bounded alarm, and a zero-cost spatial diagnostic. The method proceeds through four sequential modules, namely fluctuation extraction, sliding covariance estimation, Synchronization Index computation, and CUSUM sequential testing.

\subsection{Active-Power Fluctuation Extraction}

Raw SCADA active-power measurements at data-center substations contain two components. A slow normal load trend reflects building operation, scheduled workload changes, and cooling cycles, whereas a faster episode-driven fluctuation reflects short-duration workload activity that may become synchronized across substations. Since the detector targets the fast component, the slow trend is removed before the covariance calculation.

Each active-power channel is detrended by subtracting a causal moving average,
\begin{equation}
\Delta P_i(t) = P_i(t) - \bar{P}_i(t),
\qquad
\bar{P}_i(t) = \frac{1}{T_d}\sum_{\ell=0}^{T_d-1} P_i(t - \ell\,\Delta t),
\label{eq:detrend}
\end{equation}
where $P_i(t)$ is the raw active-power measurement at substation $i$, $\bar{P}_i(t)$ is the recent local average, $T_d$ is the detrending window, $\Delta t$ is the SCADA sampling interval, and $\ell$ indexes the $T_d$ discrete samples in the averaging window. The resulting signal $\Delta P_i(t)$ is the fast active-power fluctuation around the local operating point.

The window $T_d$ must be longer than the expected episode duration but shorter than the main load-drift timescale, so that it suppresses slow drift while preserving episode-band fluctuations. For SCADA data sampled every 1\,s and episodes lasting about 10--30\,s, $T_d = 30$ samples is a practical choice. The detrended multi-bus vector is then $\mathbf{p}(t) = [\Delta P_1(t),\ldots,\Delta P_N(t)]^\top$.

\subsection{Sliding Sample Covariance Estimation}

At each time step $t_k$, the sample covariance matrix of the detrended measurement vector is computed over a sliding window of $W$ samples,
\begin{equation}
  \hat{\mathbf{C}}(t_k) = \frac{1}{W-1}\sum_{m=0}^{W-1}
  \bigl(\mathbf{p}(t_k - m\Delta t) - \bar{\mathbf{p}}_k\bigr)
  \bigl(\mathbf{p}(t_k - m\Delta t) - \bar{\mathbf{p}}_k\bigr)^\top,
  \label{eq:covest}
\end{equation}
where $\bar{\mathbf{p}}_k = (1/W)\sum_{m=0}^{W-1}\mathbf{p}(t_k - m\Delta t)$ is the within-window mean and $m$ indexes the $W$ samples in the sliding window. The matrix $\hat{\mathbf{C}}(t_k) \in \mathbb{R}^{N \times N}$ is symmetric and positive semi-definite, and it summarizes the cross-substation statistical structure of the $W$ most recent observations. It is updated on every SCADA scan by discarding the oldest sample and adding the newest, a rank-one update that is computationally trivial. The window length $W$ governs a bias-variance trade-off. A shorter window reduces the lag between episode onset and covariance response but admits more estimation noise, whereas a longer window smooths the estimate at the cost of alarm delay. For a small cluster of $N$ substations ($N = 3$ in the case study), a window of at least $W = 20$--$30$ samples yields a well-conditioned $N \times N$ sample covariance with $W - 1$ degrees of freedom. $W = 30$\,s at SCADA rate is retained as the point balancing stable eigenvalue estimates against latency and the dilution of brief episodes. Episodes shorter than 10\,s contribute only a small fraction of the window's correlation signal at this setting, which compresses the SI elevation and limits the detection margin for those events. This effect is consistent with a detection-theoretic bound derived for correlation-based AI load classifiers~\cite{chaudhary2027gridedge}.

To make the Synchronization Index interpretable across substations with different per-bus power levels, each channel is standardized before the covariance computation. The historical standard deviation $\hat{\sigma}_i$ is estimated from a baseline period, and $\Delta P_i(t)/\hat{\sigma}_i$ replaces $\Delta P_i(t)$ in $\mathbf{p}(t)$. The resulting $\hat{\mathbf{C}}$ then approximates the correlation matrix, whose diagonal entries are unity and whose off-diagonal entries are the sliding pairwise correlations between substations.

\subsection{Synchronization Index}

The eigendecomposition of the correlation-scale covariance estimate yields $N$ eigenvalues and eigenvectors,
\begin{equation}
  \hat{\mathbf{C}}(t_k) = \sum_{j=1}^{N}\hat{\lambda}_j(t_k)\,\hat{\mathbf{v}}_j(t_k)\,\hat{\mathbf{v}}_j(t_k)^\top,
  \quad \hat{\lambda}_1 \geq \hat{\lambda}_2 \geq \cdots \geq \hat{\lambda}_N \geq 0,
  \label{eq:eig}
\end{equation}
where $j = 1,\ldots,N$ indexes the eigenvalue-eigenvector pairs in decreasing order of eigenvalue magnitude. The Synchronization Index at time $t_k$ is the fraction of total variance captured by the dominant mode~\cite{hotelling1933analysis,abdi2010principal},
\begin{equation}
  \mathrm{SI}(t_k) = \frac{\hat{\lambda}_1(t_k)}{\displaystyle\sum_{j=1}^{N}\hat{\lambda}_j(t_k)} = \frac{\hat{\lambda}_1(t_k)}{\mathrm{tr}\!\left(\hat{\mathbf{C}}(t_k)\right)},
  \label{eq:si}
\end{equation}
where $\mathrm{tr}(\cdot)$ denotes the trace of a matrix, equal to the sum of its diagonal entries and, equivalently, the sum of its eigenvalues. Three properties make SI the natural detection statistic for this problem. First, its range is analytically bounded, with $\mathrm{SI} \in [1/N, 1]$ for any $N$-bus system. The lower bound $1/N$ is achieved when all eigenvalues are equal, which occurs under full statistical independence, and is known analytically without any empirical baseline. The upper bound of 1 is approached when all of the variance is concentrated in a single eigenvector. This condition corresponds to full synchronization. Second, the SI is monotone with episode severity. As the pairwise correlations among substations increase, the dominant eigenvalue absorbs a larger fraction of total variance and SI rises toward 1, so SI is a continuous severity measure rather than a binary flag. Third, SI scales correctly with $N$, because the independence baseline is always $1/N$ regardless of how many substations are monitored. For the special case of $N$ substations with equal per-bus variance and uniform pairwise correlation $\rho$, the correlation matrix has eigenvalues $\lambda_1 = 1 + (N-1)\rho$ and $\lambda_2 = \cdots = \lambda_N = 1 - \rho$. This gives,
\begin{equation}
  \mathrm{SI} = \frac{1 + (N-1)\rho}{N},
  \label{eq:si_uniform}
\end{equation}
so $\mathrm{SI} = 1/N$ when $\rho = 0$ and $\mathrm{SI} = 1$ when $\rho = 1$. Equation~\eqref{eq:si_uniform} shows that SI is a linear function of mean pairwise correlation in this symmetric case.

\subsection{Sequential Detection via CUSUM}

The SI time series is a scalar sequence that is close to $1/N$ during normal operation and elevated during episodes. A one-sided CUSUM test~\cite{page1954cusum} converts this sequence into a sequential alarm that minimizes the expected detection delay for a specified false-alarm rate. The CUSUM accumulator is initialized to zero and updated at each SCADA scan,
\begin{equation}
  S_k = \max\bigl(0,\;S_{k-1} + \mathrm{SI}(t_k) - \mu_0 - \kappa\bigr),
  \quad S_0 = 0,
  \label{eq:cusum}
\end{equation}
where $\mu_0$ is the in-control baseline and $\kappa > 0$ is the reference value, also called the allowance, set to half the minimum detectable elevation above baseline,
\begin{equation}
  \kappa = \frac{1}{2}\,\delta, \qquad \delta = \mathrm{SI}_{\min} - \mu_0,
  \label{eq:kappa}
\end{equation}
where $\mathrm{SI}_{\min}$ is the smallest SI elevation that the operator wishes to reliably detect. An episode alarm fires at the first time step for which $S_k$ exceeds the threshold $h$. The CUSUM resets automatically to zero whenever $S_k$ would otherwise become negative, so episode end is flagged by the accumulator falling to the reset region. The threshold $h$ is set at an operating point favorable to the $F_1$ score, computed over a calibration period held out from the evaluation record.

In a system with statistically independent loads, the theoretical baseline is $\mu_0 = 1/N$, but in practice background correlations from shared infrastructure, such as cooling cycles and synchronized demand patterns, elevate the ambient SI above $1/N$. The baseline $\mu_0$ is therefore estimated as the median SI of windows identified as normal in a first analysis pass. These labels come from pairwise Pearson correlations in the covariance matrix, not from the CUSUM statistic, so no circularity arises. The resulting $\mu_0$ ensures that the CUSUM increment is negative in at least 50\% of in-control windows, so the accumulator resets frequently and false-alarm pressure remains bounded. No labeled training data is required beyond the episode-free calibration window used to estimate $\mu_0$, which can be taken as the start-up period before the AI data centers begin training workloads.

\vspace{-0.5em}
\subsection{Spatial Attribution via Participation Factors}

When the CUSUM accumulator crosses the threshold and an alarm fires, the dominant eigenvector $\hat{\mathbf{v}}_1(t_k)$ of the covariance estimate identifies how the synchronized variance is distributed across substations, a spatial counterpart to the participation factors used in power-system modal analysis~\cite{perezarriaga1982selective}. The participation factor of substation $i$ in the current episode is,
\begin{equation}
  \pi_i(t_k) = \hat{v}_{1,i}(t_k)^2, \qquad \sum_{i=1}^{N}\pi_i(t_k) = 1,
  \label{eq:participation}
\end{equation}
where $\hat{v}_{1,i}$ denotes the $i$-th component of the unit-norm eigenvector $\hat{\mathbf{v}}_1$. A substation with $\pi_i > 1/N$ contributes more than its proportional share to the synchronized mode, and a substation with $\pi_i \approx 0$ is not participating in the current episode. The participation vector is computed at every scan update, so it tracks the evolving spatial pattern as an episode develops, peaks, and decays. This output lets the operator identify and focus attention on the substations most responsible for the synchronized episode.


\section{Test System and Case Studies}\label{sec:results}

\begin{figure*}[!htbp]
  \centering
  \vspace{-1em}
  \includegraphics[width=0.9\textwidth]{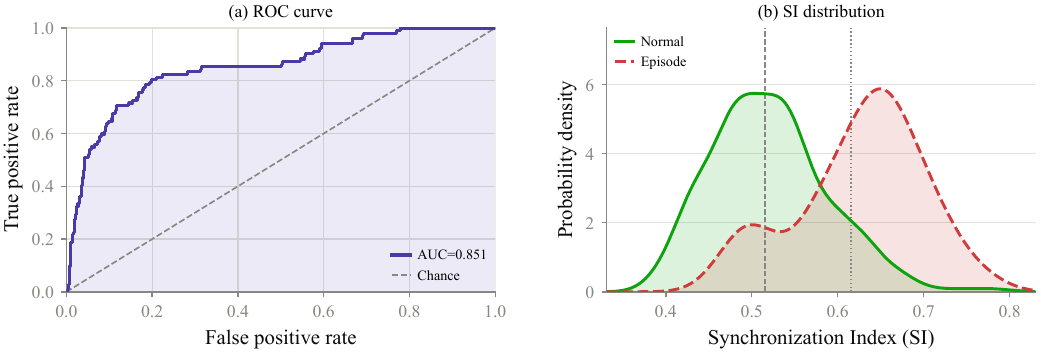}
  \vspace{-1em}
  \caption{(a) Receiver operating characteristic (ROC) curve of the SI-threshold detector on the RTDS testbed. (b) Class-conditional distribution of SI under normal ($H_0$) and episode ($H_1$) windows, with the estimated CUSUM baseline $\hat{\mu}_0$ (dashed) and detection level $\hat{\mu}_0+\delta$ (dotted) marked.}
  \vspace{-1.25em}
  \label{fig:diagnostics}
\end{figure*}

This section presents the RTDS testbed, the detector calibration, and the detection and attribution results that validate the method developed in the previous section.

\subsection{RTDS Testbed and AI Load Model}

All experiments use the IEEE New England 39-bus system simulated in the Real-Time Digital Simulator (RTDS) at Michigan State University. The testbed models all ten generator dynamics, the full network impedance matrix, and converter-coupled load interfaces at sub-millisecond time steps, so frequency and voltage responses to load changes are physically accurate. Physics-informed synthetic AI data-center loads are connected at buses~4, 12, and~15. These buses are geographically dispersed across the network, each connected to a distinct transmission voltage level, consistent with a bus-placement analysis of data-center-driven voltage and frequency stability impacts on this same test system~\cite{chaudhary2025stability}. This same three-bus configuration supports a hardware-in-the-loop edge-deployment study of the companion correlation-detection method~\cite{chaudhary2027gridedge}.

The load model at each bus follows the two-phase BSP structure described in the synchronized-episode section. A compute phase draws high, approximately constant active power, and a communication phase produces a load step driven by gradient synchronization. Phase transitions occur at $\lambda = 0.3$\,iterations per second~\cite{chaudhary2026modal,chaudhary2026semimarkov}. A 20-second thermal cycle from computer room air conditioning (CRAC) is superimposed on each bus to reproduce the building-level thermal signature documented in field measurements~\cite{epri2024powering}. The three facilities share a common orchestration platform, so their phase transitions are stochastically coupled. A synchronized episode begins when multiple buses enter their communication phase within a short time window. The simulation record spans 800\,s at 1-kHz resolution. Downsampling to 1\,s SCADA resolution yields \Nwindows{} analysis windows. The per-channel power levels at buses 4, 12, and 15 are 200--250\,MW, consistent with hyperscale data-center demand~\cite{epri2024powering}.

\subsection{Detection Method Configuration and Episode Labeling}

The detection method uses $N = 3$ buses, detrending window $T_d = 30$\,s, and covariance window $W = 30$\,s, which yields a well-conditioned estimate with $W - 1 = 29$ degrees of freedom and filters brief non-episode fluctuations, consistent with the window-length analysis above. Each channel is standardized by its baseline standard deviation estimated over the first 60\,s of the record, so that $\hat{\mathbf{C}}(t_k)$ approximates the correlation matrix and the SI baseline approaches $1/N$ in the absence of synchronization.

Background correlations driven by the shared CRAC thermal cycle raise the ambient SI to a median of approximately 0.52 across normal windows, well above the theoretical independence floor $1/N = 0.33$. The CUSUM baseline $\mu_0 = 0.516$ is estimated as the median SI of the windows labeled normal in the first analysis pass. The minimum detectable elevation is set to $\delta = 0.10$, so $\kappa = 0.05$. The alarm threshold is fixed at $h = \CUSUMh$, near the $F_1$-favorable region of the operating curve, and the mean detection latency at this operating point is $\Lmean$.

Reference episode labels are derived independently of the CUSUM detector. A SCADA window is labeled positive when the maximum pairwise Pearson correlation $\max_{i \neq j} |\hat{\rho}_{ij}|$ computed from the same sliding covariance matrix exceeds $\tau_\rho = 0.60$, a threshold consistent with prior field and simulation studies of AI training workloads~\cite{li2024unseen,chaudhary2026spatial}. Of the \Nwindows{} windows, \Npos{} (13.2\%) are labeled positive and \Nneg{} (86.8\%) are labeled normal. The 13 episode clusters in the record have durations of 1--38\,s, with a median of 5\,s. This spread reflects the stochastic coupling in the BSP orchestration model.

\subsection{Covariance Window Sensitivity}

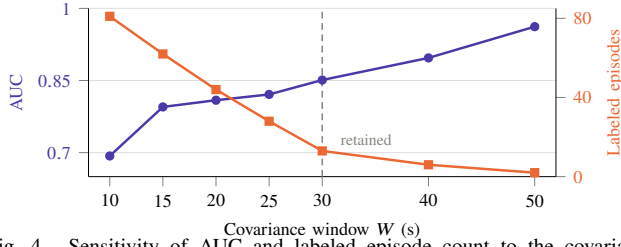
\begin{figure}[!htbp]
  \centering
  \scalebox{0.85}{%
  \begin{tikzpicture}
    \begin{axis}[
      width=\columnwidth, height=4.2cm,
      xlabel={Covariance window $W$ (s)},
      ylabel={AUC},
      xlabel style={font=\footnotesize},
      ylabel style={font=\footnotesize, color=siline},
      xmin=8, xmax=52, ymin=0.65, ymax=1.0,
      xtick={10,15,20,25,30,40,50},
      ytick={0.7,0.85,1.0},
      tick label style={font=\footnotesize},
      y tick label style={color=siline},
      tick align=outside,
      axis y line*=left,
      axis x line*=bottom,
      ymajorgrids, grid style={line width=0.2pt, draw=gray!25},
    ]
      \draw[densely dashed, refgray, line width=0.7pt] (axis cs:30,0.65) -- (axis cs:30,1.0);
      \node[font=\scriptsize, refgray, anchor=south west] at (axis cs:31,0.700) {retained};
      \addplot[color=siline, mark=*, mark size=1.5pt, line width=1.1pt]
        coordinates {(10,0.693) (15,0.795) (20,0.809) (25,0.821) (30,0.851) (40,0.897) (50,0.962)};
    \end{axis}
    \begin{axis}[
      width=\columnwidth, height=4.2cm,
      ylabel={Labeled episodes},
      ylabel style={font=\footnotesize, color=cusumline},
      xmin=8, xmax=52, ymin=0, ymax=85,
      ytick={0,40,80},
      tick label style={font=\footnotesize},
      y tick label style={color=cusumline},
      tick align=outside,
      axis y line*=right,
      axis x line=none,
    ]
      \addplot[color=cusumline, mark=square*, mark size=1.5pt, line width=1.1pt]
        coordinates {(10,81) (15,62) (20,44) (25,28) (30,13) (40,6) (50,2)};
    \end{axis}
  \end{tikzpicture}%
  }
  \vspace{-1.5em}
  \caption{Sensitivity of AUC and labeled episode count to the covariance window $W$, with the retained $W = 30$\,s operating point marked.}
  \vspace{-1.5em}
  \label{fig:window_sensitivity}
\end{figure}

The covariance window $W$ jointly sets the eigenvalue estimate used for detection and the pairwise-correlation labels used for evaluation. Its effect on both must therefore be examined together, not treated as a free parameter that maximizes a single metric in isolation. Fig.~\ref{fig:window_sensitivity} sweeps $W$ from 10\,s to 50\,s and plots the area under the ROC curve (AUC) of the SI-threshold detector alongside the number of reference episode clusters each window identifies. The AUC rises monotonically as $W$ grows, because a longer window yields a better-conditioned covariance estimate with more degrees of freedom, whereas the labeled episode count falls sharply because the wider window merges temporally adjacent episodes into a single labeled interval. At $W=50$\,s, only two labeled episodes remain, too few to support a meaningful detection-rate or latency estimate. The window itself also exceeds the 10--30\,s physical episode timescale, so the rising AUC there reflects a coarsening of evaluation granularity and not a genuine gain in detectability. $W = 30$\,s is retained as the point that keeps the window commensurate with the physical episode timescale while still yielding a well-conditioned estimate, not the value that happens to maximize AUC.

\subsection{Detection Performance}

\begin{figure}[!htbp]
  \centering
  \includegraphics[width=0.9\columnwidth]{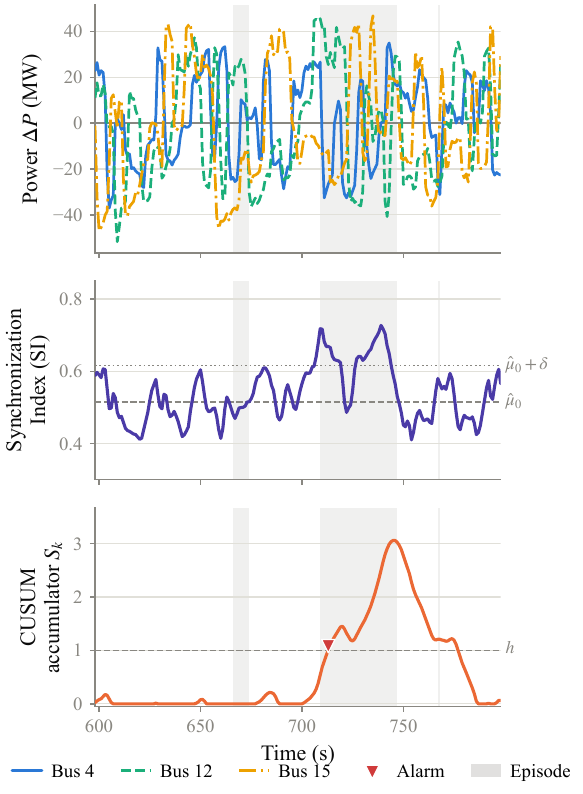}
  \vspace{-1.25em}
  \caption{Representative 200-second segment, with detrended active power (top), Synchronization Index against its baseline and detection level (middle), and CUSUM accumulator $S_k$ (bottom).}
  \vspace{-1em}
  \label{fig:si_timeseries}
\end{figure}

Table~\ref{tab:results} summarizes the detection performance. The AUC of \AUC{} confirms that the SI statistic discriminates synchronized episode windows from normal operation with a wide margin over random guessing. The detrended active power, SI time series, and CUSUM accumulator for a representative 200-second segment appear in Fig.~\ref{fig:si_timeseries}, with episode labels marked throughout. During episode windows, the SI rises above the estimated baseline $\hat{\mu}_0 = 0.516$, and the CUSUM accumulator $S_k$ responds within \Lmean{} of the first window to cross the drift level $\hat{\mu}_0 + \delta$. The detection rate of \DR{} is a window-level statistic over the \Npos{} positive windows, distinct from the cluster-level count below. The false-alarm rate of \FAR{} at $h = \CUSUMh$ follows directly from the window-length dilution established above. Only five of the 13 episode clusters are detected. The eight undetected clusters range from 3\,s to 8\,s in duration, with peak SI from 0.503 to 0.688, generally lower than the detected clusters' 0.580--0.773 range. Because $S_k$ integrates SI elevation over time instead of a single peak, a miss reflects a joint shortfall in magnitude and duration, not duration alone.

\begin{table}[!htbp]
  \centering
  \vspace{-1em}
  \caption{Detection performance of the eigenvalue-CUSUM detector}
  \label{tab:results}
  \setlength{\tabcolsep}{5pt}
  \begin{tabular}{lc}
    \toprule
    Metric & Value \\
    \midrule
    Detection Rate (DR)               & \DR \\
    False Alarm Rate (FAR)            & \FAR \\
    Area Under ROC Curve (AUC)        & \AUC \\
    Mean Detection Latency            & \Lmean \\
    \bottomrule
  \end{tabular}
  \vspace{-1.25em}
\end{table}

Fig.~\ref{fig:diagnostics}(a) and Fig.~\ref{fig:diagnostics}(b) examine the discriminative power of the SI statistic directly. The ROC curve in Fig.~\ref{fig:diagnostics}(a) traces the detection rate against the false-alarm rate as the SI threshold sweeps its full range. The curve stays well above the diagonal chance line across the entire operating range, consistent with the AUC of \AUC{} in Table~\ref{tab:results}. Fig.~\ref{fig:diagnostics}(b) shows the class-conditional distribution of SI under normal and episode windows. The two distributions separate cleanly around the estimated baseline $\hat{\mu}_0$, but they retain a region of overlap immediately above the detection level $\hat{\mu}_0 + \delta$, populated by episodes whose magnitude and duration were jointly insufficient to carry the CUSUM accumulator above threshold. This overlap is the direct cause of the false-alarm and missed-detection counts in Table~\ref{tab:results}, and it confirms that the residual error at this operating point reflects the shortfall identified above and not a weakness of the SI statistic itself.

\subsection{Spatial Attribution}

Table~\ref{tab:participation} reports the participation factor vector $[\pi_1, \pi_2, \pi_3]$ at peak severity for five labeled episodes spanning the range of peak SI observed among the 13 episode clusters, from $\mathrm{SI} = 0.503$ in Episode 1 to $\mathrm{SI} = 0.773$ in Episode 5. Only Episode 5 crosses the CUSUM alarm threshold. Episodes 1--4 illustrate the attribution mechanism on labeled episodes the detector missed. The three factors stay closest to an equal split in Episode 1, the weakest of the five, where they range only from 0.206 to 0.414. Episode 4 shows the sharpest imbalance in the set, where bus~12 carries 0.732 of the variance and bus~4 carries only 0.116. Which bus carries the smallest share varies across episodes, with bus~15 weakest in Episodes 1, 3, and 5, bus~12 weakest in Episode 2, and bus~4 weakest in Episode 4. The attribution therefore tracks the orchestration pattern specific to each episode, without pointing to a fixed weak point at one substation, at no cost beyond the eigendecomposition already performed for the SI.

\begin{table}[!htbp]
  \centering
  \vspace{-1em}
  \caption{Participation factors at peak severity for five labeled episodes.}
  \label{tab:participation}
  \begin{tabular}{ccccc}
    \toprule
    Episode & Peak SI & $\pi_1$ (Bus 4) & $\pi_2$ (Bus 12) & $\pi_3$ (Bus 15) \\
    \midrule
    1 & 0.503 & 0.414 & 0.380 & 0.206 \\
    2 & 0.586 & 0.539 & 0.210 & 0.252 \\
    3 & 0.647 & 0.237 & 0.535 & 0.228 \\
    4 & 0.688 & 0.116 & 0.732 & 0.153 \\
    5 & 0.773 & 0.392 & 0.517 & 0.091 \\
    \bottomrule
  \end{tabular}
  \vspace{-1em}
\end{table}




\section{Conclusion}\label{sec:conclusion}

This paper has developed an eigenvalue-CUSUM method that turns the structural change a synchronized episode produces in the multi-bus active-power covariance matrix into a real-time severity index and a delay-bounded alarm, using only SCADA telemetry the grid already collects. The central conclusion this work supports is that detecting a synchronized episode and attributing it to the substations that drive it requires no new instrumentation, no trained classifier, and no labeled data, because the same eigendecomposition that scores severity also yields the spatial attribution at no added cost. Validation on the RTDS testbed grounds this claim. The method separates episode and normal windows with a wide margin over chance, and the residual misses on the shortest episodes trace to the covariance window's averaging horizon, not to a weakness in the underlying statistic. For an operator, this closes a gap that single-bus threshold alarming cannot. A covariance-based statistic, computed alongside the alarms an EMS already generates, turns three isolated within-limit readings into one attributable synchronized-episode alarm.

Future work can extend this method to field SCADA records with a genuine held-out calibration split and additional substations, to substation-grade edge hardware, and to a direct quantification of the reserve adequacy impact of a detected episode in operator-actionable terms.

\section*{Acknowledgment}
This research was supported in part by the MSU Research Foundation and in part by the U.S. National Science Foundation under Grant No.~2408615.

\balance
\bibliographystyle{IEEEtran}
\bibliography{references}

\end{document}